\documentclass[%
reprint,nofootinbib,
superscriptaddress,
 amsmath,amssymb,
 aps,prd
floatfix,
]{revtex4-2}

\usepackage{graphicx}
\usepackage{dcolumn}
\usepackage{bm}

\usepackage{slashed}
\usepackage[utf8]{inputenc}
\usepackage{braket}

\usepackage[compat=1.1.0]{tikz-feynman}
\usepackage[none]{hyphenat}

\newcommand{\sint}[2]{J_{#1}^{#2}}
\newcommand{\epe}{\epsilon_e}
\newcommand{\de}{d_e}
\newcommand{\dg}{d_\gamma}
\newcommand{\ppp}{p \cdot p^\prime }

\newcommand{\bepe}{\overline{\epsilon}_e}

\begin{document}

\title{One loop reduced QED for massive fermions within an innovative formalism}

\author{Victor Miguel Banda Guzm\'an}
\email{victor.banda@upslp.edu.mx}
\affiliation{Universidad Polit\'ecnica de San Luis Potos\'i, Urbano Villal\'on 500, Col. La Ladrillera, C.P. 78363, San Luis Potos\'i, S.L.P., M\'exico}

\author{Adnan Bashir}
\email{adnan.bashir@umich.mx;abashir@jlab.org}
\affiliation{Instituto de F\'isica y Matem\'aticas,c
Universidad Michoacana de San Nicol\'as de Hidalgo, Edificio C-3, Ciudad Universitaria, C.P. 58040, Morelia, Michoac\'an, M\'exico}
\affiliation{${\dagger}$Theory Center, Jefferson Lab, Newport News, VA 23606, USA}

\author{Luis Albino}
\email{albino.fernandez@umich.mx}
\affiliation{Instituto de F\'isica y Matem\'aticas,c
Universidad Michoacana de San Nicol\'as de Hidalgo, Edificio C-3, Ciudad Universitaria, C.P. 58040, Morelia, Michoac\'an, M\'exico}

\author{Dania Rodr\'iguez-Tzintzun}
\email{1717697x@umich.mx}
\affiliation{Facultad de Ciencias F\'isico Matem\'aticas,
Universidad Michoacana de San Nicol\'as de Hidalgo, Edificio alfa, Ciudad Universitaria, 
C.P. 58040, Morelia, Michoac\'an, M\'exico}

%
%

\begin{abstract}


We carry out a detailed study of the three-point fermion-photon interaction vertex at one loop order for massive fermions in reduced quantum electrodynamics.
This calculation is carried out in arbitrary covariant gauges and space-time dimensions within a recently proposed innovative approach based upon an efficient combination of the first and second order
formalisms of quantum electrodynamics. This procedure provides a natural decomposition of the vertex into its  components which are longitudinal and transverse to the photon momentum. 
It also separates the spin and scalar degrees of freedom of a fermion interacting electromagnetically, allowing us to readily establish the gauge-independence of the Pauli form factor and compute it in an expeditious manner.
All incoming and outgoing momenta are taken off-shell at the outset. However, we  present results for cases of particular kinematic interest whenever required. For the sake of completeness, we also provide expressions for
the massive fermion self energy and photon vacuum polarization, verifying known expressions for massless reduced quantum electrodynamics and computing the renormalization constants ${\cal Z}_1$, ${\cal Z}_2$ and ${\cal Z}_3$. As we provide general expressions for the computed Green functions, we readily reproduce and confirm the results for standard quantum electrodynamics. Comparing the two cases, we infer that the Pauli form factor for reduced quantum electrodynamics is $8/3$ times that for the standard QED in four dimensions, implying a higher Land\'e $g$-factor. We expect our perturbative calculation of the fermion-photon vertex to serve as a guide for any non-perturbative construction of this Green function, invariably required in the Schwinger-Dyson equation studies of the subject. We also comment on the Landau-Khalatnikov-Fradkin transformations of the massive fermion propagator and provide a comparison with its one loop calculation.
\end{abstract}

\maketitle

\section{Introduction}

Electromagnetism is a fundamental force of nature, more studied and far better understood than any other interaction at the elementary level, i.e., strong or weak interactions. It is characterized by the following action of quantum electrodynamics (QED)~: 
\begin{eqnarray}
    && \hspace{-5mm}   S = \int d^{\de}x \Big[ -\dfrac{1}{4} F_{\mu\nu} F^{\mu\nu} - \dfrac{1}{2 \xi} ( \partial_\mu A^\mu )^2 + \overline{\psi}( i \slashed{\partial} - m )\psi \nonumber \\
 && \hspace{15mm}  - \,  e \overline{\psi} \gamma^\mu \psi A_\mu \Big] \,, \label{standard_QED_action}
\end{eqnarray}
where $\psi$ is the fermion field, $A^\mu$ is the gauge field, $F^{\mu\nu}$ is the electromagnetic tensor field, $e$ is the electric charge of the fermion, $\xi$ is the covariant gauge fixing parameter and $\de$ is the number of space-time dimensions (see appendix \ref{conventions} for our conventions). For $\de=4$, the quantum field theory associated with the action of Eq.~\eqref{standard_QED_action} is renormalizable, while for $\de=3$ it is super-renormalizable. Though QED has been extensively studied for more than 50 years, it continues to be an active field of research. For example, precision tests of the standard model, see for example~\cite{Aoyama:2020ynm,Colangelo:2022jxc}, and the use of electromagnetic probes to unravel the internal of structure of hadrons, e.g.~\cite{Aznauryan:2012ba,Aguilar:2019teb,Hernandez-Pinto:2023yin}, ensure QED interactions are increasingly relevant and important. 

Moreover, new variations of standard QED find applications in condensed matter systems. In the last decades, for example, a considerable research effort has been dedicated to study an extended version of the action in Eq.~\eqref{standard_QED_action} which includes non-local operators~\cite{Gorbar_RQED, Teber_2018, Marino_pseudoQED, Nonlocal_QED, Marino_unitarity} (for scalar quantum field theories with non local operators see~\cite{Nonlocal_scalar_QFT}). Such a QED action $S_{\text{NL}}$,  where NL stands for non-local, can be written as~\cite{Nonlocal_QED},
\begin{eqnarray}
 &&  \hspace{-5mm} S_{\text{NL}} = \int d^{\de}x \Big[ -\dfrac{1}{4} F_{\mu\nu} D^{s-2} F^{\mu\nu} - \dfrac{1}{2 \xi}  \partial_\mu A^\mu D^{s-2} \partial_\nu A^\nu  \nonumber \\ && \hspace{19mm} + \, \overline{\psi}( i \slashed{\partial} - m )\psi - e \overline{\psi} \gamma^\mu \psi A_\mu \Big] \,, \label{Nonlocal_QED_action} 
\end{eqnarray}
where the non-local derivative $D^s$ is defined as:
\begin{eqnarray}
    \int d^{\de}x\; D^s \phi(x) e^{ikx} = |k|^s \hat{\phi}(k) \,,
\end{eqnarray}
with $s$ being a real number and
\begin{eqnarray}
     \hat{\phi}(k) = \int d^{\de}x \; \phi(x) e^{ikx} \,.
\end{eqnarray}
As stated before, in contrast with a mere academic interest, these theories, dubbed in the literature as pseudo-QED or  reduced QED (RQED), have potential applications in condensed matter physics, particularly to investigate the properties of graphene (see for example~\cite{RQED_app1, RQED_g2, RQED_curved}). The name carries the adjective {\em reduced} because some of the non local quantum field theories can be equivalently described by a dimensional reduction of a bulk gauge field in $d_\gamma$ dimensions interacting with a fermionic field in $d_e$ dimensions as discussed in the Refs.~\cite{Marino_pseudoQED, Teber2012, Teber_2018, Nonlocal_QED}. The corresponding action, $S_{\text{RQED}}$, can be written as:
\begin{eqnarray}
&&  \hspace{-5mm}  S_{\text{RQED}} = \int d^{d_\gamma}x \Big[ -\dfrac{1}{4} F_{mn} F^{mn} - \dfrac{1}{2 \xi} ( \partial_m A^m )^2 \Big] \nonumber \\
&&  \hspace{7mm} + \int d^{d_e}x \Big[ \overline{\psi}( i \slashed{\partial} - m )\psi - e \overline{\psi} \gamma^\mu \psi A_\mu \Big] \,, \label{reduced_QED_action} 
\end{eqnarray}
where the Roman indices $m,n$ associated with the bulk coordinates run from $0, \dots, d_{\gamma}-1$, while the Greek indices associated with the fermion coordinates run from $0, \dots, d_e-1$. Here one assumes that $\dg \geq \de$, and that the gauge field in the interaction term is evaluated at $A^\mu(x_0, \dots, x_{\de}, 0, \dots, 0)$.

For the action defined in Eq.~\eqref{reduced_QED_action}, the corresponding generating function $Z(\eta,\overline{\eta},J^m)$ for RQED  reads as
\begin{eqnarray}
        && \hspace{-9mm} Z(\eta,\overline{\eta},J^m) = N \; \text{Exp}\Bigg[ -ie \int d^{\dg}x\, \delta(\overline{x}) \delta^{m\mu} \nonumber \\
        && \times \left( \dfrac{1}{i}\dfrac{\delta}{\delta J^m(x)} \right) \left( i\dfrac{\delta}{\delta \eta(x_e)} \right) \gamma_{\mu} \left( \dfrac{1}{i}\dfrac{\delta}{\delta \overline{\eta}(x_e)} \right)\Bigg] Z_0 \,, \label{Z_RQED}
\end{eqnarray}
where
\begin{eqnarray}
\delta(\overline{x}) &=& \delta(x_{\de+1})\dots\delta(x_{\dg}) \,, \nonumber \\
\delta^{m\mu} &=& \begin{cases}
        1 & \text{if } m=\mu \,, \quad \text{for } \mu=0,\dots,\de -1  \\
        0 & \text{otherwise} \,,
    \end{cases}
\end{eqnarray}
$x_e$ is a short-hand notation for $x_0,\dots,x_{\de}$, $N$ is a normalization constant such that $Z(0,0,0)=1$, and $Z_0$ is defined as
\begin{eqnarray}
 && \hspace{-10mm}   Z_0 = \text{Exp}\Bigg[ -i \int d^{\de}x d^{\de}y \, \overline{\eta}(x)S(x,y)\eta(y) \nonumber \\
    && \hspace{8mm} + \dfrac{i}{2} \int d^{\dg}x d^{\dg}y \, J^m(x)\Delta_{mn}(x,y)J^n(y) \Bigg],
\end{eqnarray}
where
\begin{eqnarray}
   && \hspace{-9mm} S(x,y) = \hspace{-1mm} \int \hspace{-1mm}  \dfrac{d^{\de}p}{(2\pi)^{\de}}\, e^{-ip(x-y)}\, \dfrac{\slashed{p}+m}{p^2-m^2}, \nonumber \\
    && \hspace{-9mm} \Delta_{mn}(x,y)  = 
     \hspace{-1mm} \int \hspace{-1mm} \dfrac{d^{\dg}k}{(2\pi)^{\dg}} \dfrac{e^{-ik(x-y)}}{k^2} \Big( \hspace{-1mm} \eta_{mn}  -(1-\xi) \dfrac{k_m k_n}{k^2} \hspace{-1mm} \Big).
\end{eqnarray}
According to the generating function of Eq.~\eqref{Z_RQED}, the external legs in Feynman diagrams are amputated as in standard QED with on-shell photons represented by the polarization vectors $\epsilon^\mu_\lambda(k)$, but with the momentum variables for both photons and fermions restricted to the reduced space with dimensions $\de$. When the internal photon appears in coordinate space, we need to perform the following type of integrals,
\begin{eqnarray}
    && \int d^{\dg} x\, d^{\dg}y\, \delta(\overline{x}) \delta(\overline{y})\, \delta_{\mu m} \delta_{\nu n}\, \Delta_{mn}(x,y) \nonumber \\
    &=&  \int d^{\de}x d^{\de}y \int \dfrac{d^{\de}k}{(2\pi)^{\de}} \, e^{-ik(x-y)} \, \int \dfrac{d^{\dg-\de}\overline{k}}{(2\pi)^{\dg-\de}}  \nonumber \\
    && \times \dfrac{1}{k^2 - \overline{k}^2} \left( \eta_{\mu\nu} +(1-\xi) k_\mu k_\nu  \right),
\end{eqnarray}
where $\overline{k}^2 = k^2_{\de+1}+\dots+k^2_{\dg}$.
After integrating over the $\overline{k}$ variable, the integrals above can be rewritten as
\begin{eqnarray}
 && \hspace{-10mm}  \int d^{\dg} x\, d^{\dg}y\, \delta(\overline{x}) \delta(\overline{y})\, \delta_{\mu m} \delta_{\nu n}\, \Delta_{mn}(x,y) \nonumber \\
    &&  \hspace{-5mm}  =  - \int d^{\de}x d^{\de}y\, \int \dfrac{d^{\de}k}{(2\pi)^{\de}} \, e^{-ik(x-y)} \,\tilde{\Delta}_{\mu\nu}(k) \,,
\end{eqnarray}
where 
\begin{eqnarray}
    \tilde{\Delta}_{\mu\nu}(k) = \dfrac{1}{(4\pi)^{\epe}} \, \dfrac{\Gamma(1-\epe)}{(-k^2)^{1-\epe}} \left( \eta_{\mu\nu} -(1-\tilde{\xi}) \dfrac{k_\mu k_\nu}{k^2} \right) \,, \nonumber \\ \label{Delta_t}
\end{eqnarray}
with 
\begin{eqnarray}
    \epe = \dfrac{\dg-\de}{2} \,, \qquad \tilde{\xi} = \epe + (1-\epe)\xi \,. \label{epe_xitilde_def}
\end{eqnarray}
Thus, the integration on the coordinate space variables that do not participate in the interaction of the bulk gauge field with the fermion field produces the effective photon propagator defined in Eq.~\eqref{Delta_t}. It is the same type of propagator that arises from the non-local action of Eq.~\eqref{Nonlocal_QED_action}, (see for example~\cite{Nonlocal_QED}). 

\begin{figure}[h!]
\begin{center}
 \vspace{-1cm}
\includegraphics[scale=1]{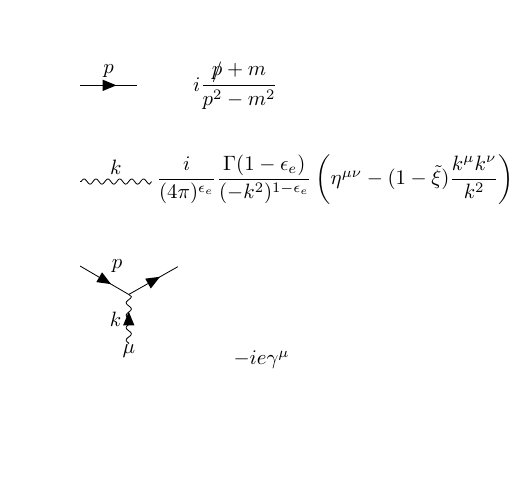}
\end{center}
\vspace{-2.1cm}
\caption{Feynman rules for RQED.} \label{Feynman_rules_RQED}
\end{figure}

According to the discussion above, the Feynman rules for the action given by Eq.~\eqref{reduced_QED_action} correspond to the ones displayed in Fig.~\ref{Feynman_rules_RQED}, which are merely the massive version of the Feynman rules derived in 
Ref.~\cite{Teber_2018}. 

In the case where $\dg=\de$, we have $\epe=0$, yielding the usual Feynman rules for standard QED. To the best of our knowledge, there is no in-depth analysis of the one loop massive RQED theories though there are related works for the massless case. Moreover, as we pointed out in our recent work~\cite{Guzman:2023hzq}, most computations of one loop fermion-boson vertex are carried out through employing the standard first order formalism,
         see for example,~\cite{Ball:1980ay,Curtis:1990zs,Kizilersu:1995iz,Bashir:1999bd,Bashir:2000rv,Bashir:2001vi,Kizilersu:2009kg,Bashir:2011dp,Bermudez:2017bpx,Albino:2022efn}.
 In this article, following the procedure detailed in~\cite{Guzman:2023hzq}, we combine this approach with the second order formalism, Refs.~\cite{Hostler,Morgan,WL_fprop1,WL_fprop2}. 
We primarily compute the three-point vertex in terms of scalar integrals using the Feynman rules for massive RQED, Fig.~\ref{Feynman_rules_RQED} and invoking this innovative formalism. Consequently, we evaluate Pauli form factor for RQED and compare it with standard QED result in $4$ dimensions. For the sake of completeness, we also present results for one loop fermion and photon propagators for massive RQED
and compute the renormalization constants ${\cal{Z}}_1$, ${\cal{Z}}_2$ and ${\cal{Z}}_3$. 

The article is organized as follows: In Sec.~\ref{Vertex_function} we generalize the results of Ref.~\cite{Guzman:2023hzq} for standard QED to decompose the fermion-photon vertex function $V^\mu$ of RQED at one loop into longitudinal and transverse components in arbitrary gauge and dimensions. We express these components in terms of scalar Feynman integrals. In Sec.~\ref{Form_factors}, we use the results of the previous section to obtain the Dirac and Pauli form factors for a fermion in RQED. The expressions obtained for these form factors are also evaluated for some particular kinematic cases of interest. In Sec.~\ref{fermion_selfEnergy}, we obtain a general expression for the fermion self energy $\Sigma$ and photon vacuum polarization $\pi(p^2)$ of RQED in terms of Feynman scalar integrals. 
From the results obtained for the vertex function, the fermion self energy and the photon vacuum polarization,  we identify ultraviolet divergent integrals when $\dg=\de=4$, and when $\dg=4$ with $\de=3$ in Sec.~\ref{RenormalizationConstants}. Once we identify these Feynman integrals, we compute the renormalization constants ${\cal{Z}}_1$, ${\cal{Z}}_2$ and ${\cal{Z}}_3$. Concluding remarks and a summary of the main results of the article are provided in Sec.~\ref{conclusions}. The manuscript is complemented with three appendices with supplemental information. Appendices \ref{scalar_integrals} and \ref{scalar_integrals_2} contain useful identities for the three-point on-shell and the two-point scalar integrals that appear in the computation of the Dirac and Pauli form factors, fermion self energy and the photon vacuum polarization. 
Appendix D contains discussion on the Landau-Khalatnikov-Fradkin (LKF) transformations for massive fermion propagator.


\section{The one loop Vertex $V^\mu(p',p)$} \label{Vertex_function}

\begin{figure}[h!]
\begin{center}
 \vspace{-2cm}
\includegraphics[scale=1]{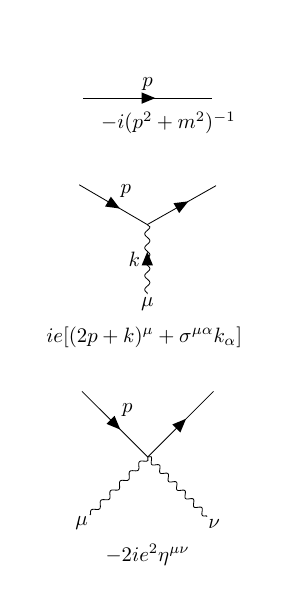}
\end{center}
\vspace{-1.2cm}
\caption{Second order rules for spinor QED.} \label{SecondOR}
\end{figure}

While we refer the reader to Ref.~\cite{Guzman:2023hzq} for all relevant details and the nitty-gritty of the combined first and second order formalisms, it might be important to flash the Feynman rules for the second order formalism, see Fig.~\ref{SecondOR}. It is relevant to the discussion on the one loop vertex and the efficient identification of the Pauli form factor and its gauge independence. Consequently,
 \begin{enumerate}
 \item 
 We arrive at our result efficiently and the decomposition of the fermion-photon vertex into its longitudinal and transverse components is achieved naturally without a Ball-Chiu~\cite{Ball:1980ay} decomposition.

\item 

Moreover, the combined analysis allows us to track those terms  which identically vanish when external momenta are taken on-shell.
The operator $(\slashed{p}-m)$ remains on the far right whereas $(\slashed{p}+m)$ is kept on the left. Therefore, when on-shell conditions $(\slashed{p}-m)u_s(p)=0$ and $\overline{u}_s(p)(\slashed{p}+m)=0$ are imposed, these terms are identically zero. 
 Employing on-shell  symmetries of the Feynman integrals, we observe that the terms which depend explicitly on the covariant gauge parameter $\xi$ in the evaluation of the on-shell Pauli form factor vanish. These cancellations lead to a compact expression for this quantity in an arbitrary space-time dimension in terms of scalar integrals with: (i) higher powers of scalar propagators and (ii) shifted dimensions. It hints towards a possible analogous simplification in the evaluation of the anomalous magnetic moment of charged fermions at higher orders of perturbation theory. 
 
\end{enumerate}

\begin{figure}[h!]
\begin{center}
 \vspace{-2cm}
\includegraphics[scale=1]{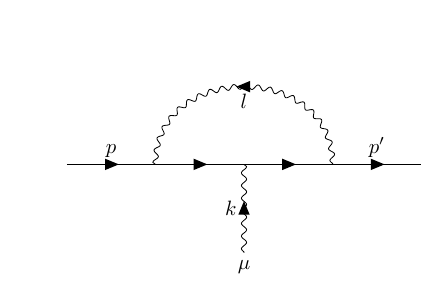}
\end{center}
\vspace{-1cm}
\caption{One loop Feynman diagram for the vertex function $V^\mu(p',p)$.} \label{Vertex}
\end{figure}

\begin{widetext}

Fig.~\ref{Vertex} depicts Feynman diagram for the one loop vertex function $V^\mu$ at order $e^3$. According to the Feynman rules given in Fig.~\ref{Feynman_rules_RQED}, the vertex function reads as follows:
\begin{eqnarray}
- i V^\mu(p',p) &=& \dfrac{e^3 \Gamma(1-\epsilon_e)}{(4\pi)^{\epsilon_e}} \int \dfrac{d^{d_e}l}{(2\pi)^{d_e}}
\gamma^\rho \, \dfrac{\slashed{p}^\prime+\slashed{l}+m}{(p^\prime+l)^2-m^2} \gamma^\mu \, \dfrac{\slashed{p}+\slashed{l}+m}{(p+l)^2-m^2}\, \dfrac{\gamma^\nu}{(-l^2)^{1-\epsilon_e}} \left(\eta_{\rho\nu}-(1-\tilde{\xi}) \dfrac{l_\rho l_\nu}{l^2}\right) \,. \label{V_1L}
\end{eqnarray}
To evaluate the one loop expression of the vertex given in  Eq.~\eqref{V_1L}, we follow the approach presented in Ref.~\cite{Guzman:2023hzq}. There, the Feynman rules of the first order formalism, Fig.~\ref{Feynman_rules_RQED}, were merged with those of the corresponding second order formalism, Fig.~\ref{SecondOR}, to obtain a natural decomposition of the vertex into longitudinal and transverse components with respect to the external photon momentum $k^\mu$. We generalize the identities employed in  Ref.~\cite{Guzman:2023hzq} to arbitrary power of the internal photon propagator and implement them to decompose and evaluate Eq.~\eqref{V_1L} as follows:
\begin{eqnarray}
    V^\mu = V_L^\mu + V_T^\mu \,, \label{VRQED}
\end{eqnarray}
where the longitudinal $V_L^\mu$ and transverse $V_T^\mu$ vertex components can be written in an apparently cumbersome yet logically straightforward notation:
\begin{eqnarray}
V_L^\mu &=& \dfrac{e^3\Gamma(\bepe)}{(4\pi)^{\epe+\frac{\de}{2}}} \Big\{ m(1-\tilde{\xi}-\de)(\sint{1,1,\bepe}{\de}-2\sint{1,2,\bepe}{\de+2})p^\mu + m(1-\tilde{\xi}-\de)(\sint{1,1,\bepe}{\de}-2\sint{2,1,\bepe}{\de+2})p^{\prime \mu} + \Big[ (2-\de) \sint{1,1,\bepe}{\de+2} \nonumber \\
&& + (\de-3+\tilde{\xi}) \sint{1,0,\bepe}{\de} + (m^2+p^2)(1-\tilde{\xi})\sint{1,1,1+\bepe}{\de+2} \Big] \gamma^\mu - \Big[ (2-\de)\left( \sint{1,1,\bepe}{\de} - 3 \sint{1,2,\bepe}{\de+2} + 4 \sint{1,3,\bepe}{\de+4} \right) \nonumber \\
&& + (1-\tilde{\xi})\big( \sint{1,1,\bepe}{\de} - \sint{1,1,1+\bepe}{\de+2} - \sint{1,2,\bepe}{\de+2} + 4 m^2 \sint{1,3,1+\bepe}{\de+4} + 2 p \cdot p^\prime \sint{2,2,1+\bepe}{\de+4} \big) \Big] p^\mu \slashed{p}  - \Big[ (\de-2)\left( \sint{2,1,\bepe}{\de+2} - 2 \sint{2,2,\bepe}{\de+4} \right) \nonumber \\
&& + (1-\tilde{\xi})\big( \sint{2,1,\bepe}{\de+2} + 2m^2 \sint{2,2,1+\bepe}{\de+4} + 4 p \cdot p^\prime \sint{3,1,1+\bepe}{\de+4} \big) \Big] p^\mu \slashed{p}^\prime - \Big[ (2-\de)\big( \sint{1,1,\bepe}{\de} - \sint{1,2,\bepe}{\de+2} - 2 \sint{2,1,\bepe}{\de+2} + 2 \sint{2,2,\bepe}{\de+4}\big) \nonumber \\
&& + (1-\tilde{\xi}) \Big( \sint{1,1,\bepe}{\de} - \sint{1,1,1+\bepe}{\de+2} + \sint{1,2,\bepe}{\de+2} + 4 p^2 \sint{1,3,1+\bepe}{\de+4}  - 2 \sint{2,1,\bepe}{\de+2} + 2 (m^2-p^2+p \cdot p^\prime) \sint{2,2,1+\bepe}{\de+4} \big) \Big] p^{\prime \mu } \slashed{p} \nonumber \\
&& - \Big[ (\de-2) \left( \sint{2,1,\bepe}{\de+2}-4\sint{3,1,\bepe}{\de+4} \right) + (1-\tilde{\xi})\Big( \sint{2,1,\bepe}{\de+2} +2p^2 \sint{2,2,1+\bepe}{\de+4} + 4(m^2-p^2+p\cdot p^\prime)\sint{3,1,1+\bepe}{\de+4} \Big) \Big] p^{\prime \mu} \slashed{p}^\prime \Big\} \,,  \label{VL} \\
\nonumber \\
V_T^\mu &=& \dfrac{e^3 \Gamma(\bepe)}{(4\pi)^{\epe+\frac{\de}{2}}} \Big\{ \Big( 2 (\tilde{\xi}-1) \Big[ \sint{2,2,1+\bepe}{\de+4} \left( p^\mu \slashed{k} - p \cdot k \gamma^\mu \right) \slashed{p}^\prime + \sint{2,2,1+\bepe}{\de+4} \left( p^{\prime \mu} \slashed{k} - p^\prime \cdot k \gamma^\mu  \right) \slashed{p} + 2 \sint{1,3,1+\bepe}{\de+4} \left( p^\mu \slashed{k} - p \cdot k \gamma^\mu \right) \slashed{p} \nonumber \\
&& + 2 \sint{3,1,1+\bepe}{\de+4} \left( p^{\prime \mu} \slashed{k} - p^\prime \cdot k \gamma^\mu  \right) \slashed{p}^\prime \Big] + \Big[ (4-\de) \sint{1,1,\bepe}{\de} + (1-\tilde{\xi})(\sint{1,1,\bepe}{\de}-2 \sint{1,1,1+\bepe}{\de+2}) \Big] \sigma^{\mu\alpha} k_\alpha  \Big) (-\slashed{p}+m) \nonumber \\
&& + \Big[ 2 \sint{1,1,\bepe}{\de} + (\de-6) \sint{1,2,\bepe}{\de+2} -(1-\tilde{\xi})\Big( \sint{1,1,1+\bepe}{\de+2} - 2 p \cdot p^\prime \sint{2,2,1+\bepe}{\de+4} - 4 p^2 \sint{1,3,1+\bepe}{\de+4} - \sint{1,2,\bepe}{\de+2} \Big) \Big] \sigma^{\mu \alpha}k_\alpha \slashed{p} \nonumber \\
&& - \Big[ (6-\de) \sint{2,1,\bepe}{\de+2} - (1-\tilde{\xi})\Big( 2p^2 \sint{2,2,1+\bepe}{\de+4} + 4 \ppp \sint{3,1,1+\bepe}{\de+4} + \sint{2,1,\bepe}{\de+2} \Big) \Big] \sigma^{\mu \alpha }k_\alpha \slashed{p}^\prime  \nonumber \\
&& + 2 \Big[ 2\sint{1,1,\bepe}{\de} - 2 \sint{1,2,\bepe}{\de+2} - (1-\tilde{\xi})\Big( \sint{1,1,1+\bepe}{\de+2} - 2 \ppp \sint{2,2,1+\bepe}{\de+4}-4p^2 \sint{1,3,1+\bepe}{\de+4}-\sint{1,2,\bepe}{\de+2} \Big) \Big](p^\mu \slashed{k}-k \cdot p \gamma^\mu)  \nonumber \\
&& - 2 \Big[ 2 \sint{2,1,\bepe}{\de+2} - (1-\tilde{\xi})\Big( 2p^2 \sint{2,2,1+\bepe}{\de+4} + 4 \ppp \sint{3,1,1+\bepe}{\de+4} + \sint{2,1,\bepe}{\de+2} \Big) \Big](p^{\prime \mu}\slashed{k} -p^\prime \cdot k \gamma^\mu ) \Big\} \,, \label{VT}
\end{eqnarray}
where $\sigma^{\mu\nu} = \frac{1}{2}[\gamma^\mu,\gamma^\nu]$, $\bepe=1-\epe$, and the scalar integrals $\sint{a,b,c}{D}$ are defined as
\begin{eqnarray}
J_{a,b,c}^{D}(p,p^\prime) = \int \dfrac{d^{D}l}{i\pi^{\frac{D}{2}}} \dfrac{1}{\left[ -(p^\prime+l)^2+m^2 \right]^a\left[ -(p+l)^2+m^2 \right]^b(-l^2)^c} \,. \label{scalar_int}
\end{eqnarray}
The expressions obtained thus far consider all external momenta  off-shell. In order to derive Dirac and Pauli form factors, we go on-shell in the next section.

\section{Dirac and Pauli Form factors } \label{Form_factors}

Evaluation of explicit expressions for Dirac and Pauli form factors requires imposing on-shell conditions for the incoming and outgoing fermions:
\begin{eqnarray}
(\slashed{p}-m) u_s(p) = 0 \,, \qquad \overline{u}_{s^\prime}(p^\prime) (\slashed{p}^\prime - m ) = 0 \,.
\end{eqnarray}
We keep the photon off-shell for now. The spinor bi-linear covariants  $\overline{u}_s(p^\prime) V_{L,T}^\mu u_s(p)$, with $V_{L,T}^\mu$ given by  Eqs.~\eqref{VL} and \eqref{VT} thus acquire the following form:
\begin{eqnarray}
\overline{u}_{s^\prime}(p^\prime) V^{\mu}_{L}  u_s(p) &=& \dfrac{e^3 \Gamma(\bepe)}{(4\pi)^{\epsilon_e+\frac{d_e}{2}}} \Big\{ \left[ (d_e+\tilde{\xi}-3) J_{1,0,\bepe} + (2-d_e) J_{1,1,\bepe}^{d_e+2} + 2m\, f_{2L} \right]\gamma^\mu + f_{2L} \sigma^{\mu\nu}k_\nu \Big\} \,, \label{VL_OS} \\
\overline{u}_{s^\prime}(p^\prime) V^{\mu}_{T}  u_s(p) &=& \dfrac{e^3 \Gamma(\bepe)}{(4\pi)^{\epsilon_e+\frac{d_e}{2}}} \bigg( f_{1T}\, \gamma^\mu - f_{2T}\, \sigma^{\mu\nu}k_\nu \bigg) \,, \label{VT_OS}
\end{eqnarray}
where we have used the on-shell symmetry of the scalar integrals $\sint{a,b,c}{D}=\sint{b,a,c}{D}$, and applied the well-known Gordon identity
\begin{eqnarray}
\overline{u}_{s^\prime}(p^\prime) \left[ (p+p^\prime)^\mu - \sigma^{\mu\nu}k_\nu \right]  u_s(p) = 2m \overline{u}_{s^\prime}(p^\prime) \gamma^\mu  u_s(p) \,. 
\end{eqnarray}
The conveniently defined scalar functions $f_{2L}$, $f_{1T}$, $f_{2T}$ that appear explicitly in Eqs.~\eqref{VL_OS} and~\eqref{VT_OS} can be readily identified as  
\begin{eqnarray}
f_{2L} &=& -2m\, J_{1,1,\bepe}^{d_e} + 4m\, (d_e-2) J_{3,1,\bepe}^{d_e+4} + 2m(4-d_e) J_{2,1,\bepe}^{d_e+2} - 2m(2-d_e) J_{2,2,\bepe}^{d_e+4} \nonumber \\
&& + m (1-\tilde{\xi}) \Big[ J_{1,1,1+\bepe}^{d_e+2} - 2(m^2+p\cdot p^\prime) J_{2,2,1+\bepe}^{d_e+4}-4(m^2+p \cdot p^\prime) J_{3,1,1+\bepe}^{d_e+4} - 2 J_{2,1,\bepe}^{d_e+2} \Big] \,, \nonumber \\
f_{1T} &=& -2(4-d_e) k \cdot p^\prime\, J_{2,1,\bepe}^{d_e+2} - 4 k \cdot p\,J_{1,1,\bepe}^{d_e} + 4 k \cdot p\,J_{2,1,\bepe}^{d_e+2} \nonumber \\
&& + 2 k\cdot p (1-\tilde{\xi})  \Big[]   J_{1,1,1+\bepe}^{d_e+2}  - 2 p \cdot p^\prime J_{2,2,1+\bepe}^{d_e+4} - 4m^2\, J_{3,1,1+\bepe}^{d_e+4} - J_{2,1,\bepe}^{d_e+2} \Big], \nonumber \\
f_{2T} &=& 2 m(6-d_e)\,J_{2,1,\bepe}^{d_e+2} - 2m\,J_{1,1,\bepe}^{d_e} \nonumber \\
&& + m (1-\tilde{\xi}) \Big[ J_{1,1,1+\bepe}^{d_e+2} - 2(m^2+p\cdot p^\prime) J_{2,2,1+\bepe}^{d_e+4}  -4(m^2+p \cdot p^\prime) J_{3,1,1+\bepe}^{d_e+4} - 2 J_{2,1,\bepe}^{d_e+2} \Big] \,.
\end{eqnarray}
The spinor product $\overline{u}_s(p^\prime) V^\mu u_s(p)$ can thus be written in the standard form
\begin{eqnarray}
\overline{u}_{s^\prime}(p^\prime) V^{\mu}  u_s(p) = e \overline{u}_{s^\prime}(p^\prime) \left[ F_1(k^2)\, \gamma^\mu - \dfrac{1}{2m} F_2(k^2) \sigma^{\mu\nu} k_\nu \right]  u_s(p) \,,
\end{eqnarray}
where, explicitly, 
\begin{eqnarray}
F_1(k^2) &=& \dfrac{e^2 \Gamma(\bepe)}{(4\pi)^{\epsilon_e+\frac{d_e}{2}}} \Big\{ (\de-2)\sint{1,0,\bepe}{\de} + (2-\de)\sint{1,1,\bepe}{\de+2} -2(2m^2-k^2)\sint{1,1,\bepe}{\de} \nonumber \\
&& + \left[ 4(4-\de)m^2-(6-\de)k^2 \right]\sint{2,1,\bepe}{\de + 2} + 4m^2(\de-2)\left( \sint{2,2,\bepe}{\de+4}+2\sint{3,1,\bepe}{\de+4} \right) \nonumber \\
&& -(1-\tilde{\xi})\Big[ \sint{1,0,\bepe}{\de} - (2m^2-k^2)\sint{1,1,1+\bepe}{\de+2} + (4m^4+(2m^2-k^2)^2)\sint{2,2,1+\bepe}{\de+4} \nonumber \\
&& +8m^2(2m^2-k^2)\sint{3,1,1+\bepe}{\de+4} + (4m^2-k^2)\sint{2,1,\bepe}{\de+2} \Big] \, \Big\} \,, \label{F1} \\
F_2(k^2) &=& \dfrac{4e^2 m^2 \Gamma(\bepe)}{(4\pi)^{\epsilon_e+\frac{d_e}{2}}} \left[ 2 J_{2,1,\bepe}^{d_e+2} + (2-d_e)( 2 J_{3,1,\bepe}^{d_e+4} + J_{2,2,\bepe}^{d_e+4} ) \right] \,. \label{F2}
\end{eqnarray}
These are the general expressions for the Dirac and Pauli form factors at one loop order for RQED as a funcion of the photon momentum squared $k^2$ while the incoming and outgoing fermions are on-shell. 
We now discuss each of these form factors in greater detail in the following subsections.

\subsection{The Dirac form factor}

Focusing on the the Dirac form factor $F_1(k^2)$ given in Eq.~\eqref{F1}, we can mould it into a compact form if we make systematic use of the equations 
in~\eqref{sint1}. It can be rewritten as a linear combination of the scalar integrals $\sint{0,2,\bepe}{\de}$, $\sint{1,1,\bepe}{\de}$ and $\sint{1,0,\bepe}{\de}$ as follows:
\begin{eqnarray}
F_1(k^2) &=& \dfrac{e^2 \Gamma(1-\epsilon_e)}{(4\pi)^{\epsilon_e+\frac{d_e}{2}}} 
 \Bigg\{ \left( \dfrac{1}{2(\de-2+\epe)(\de-3+\epe)} \right) \Bigg[ 2(\de-2)(3-\de-\epe)^2 \sint{1,0,\bepe}{\de} \nonumber \\
&& + \Big( 4m^2[ 4-2\de +(2-4\de + \de^2)\epe + 2(\de-1) \epe^2 ] + k^2 [ \de^3 + 3\de^2(\epe-3) \nonumber \\
&& -8(\epe-2)^2 + 2\de(15-10\epe + \epe^2) ] \Big) \sint{1,1,\bepe}{\de} \Bigg] - (1-\tilde{\xi}) \Bigg( \sint{1,0,\bepe}{\de} + \dfrac{2m^2\sint{2,0,\bepe}{\de}}{\de-3+\epe} \Bigg) \Bigg\} \,. \label{F1_1}
\end{eqnarray}
Since
\begin{eqnarray}
    \sint{a,0,b}{\de} &=&  \dfrac{ \Gamma(\de-a-2b)}{\Gamma(\de-a-b)\Gamma(a)} \Gamma\left( a+b-\dfrac{\de}{2} \right)\, (m^2)^{\frac{\de}{2}-a-b} \,,
\end{eqnarray}
we readily obtain
\begin{eqnarray}
    \sint{1,0,\bepe}{\de} + \dfrac{2m^2\sint{2,0,\bepe}{\de}}{\de-3+\epe} = 0 \,.
\end{eqnarray}
Thus the Dirac form factor in 
Eq.~\eqref{F1_1} reduces to a gauge-independent expression:
\begin{eqnarray}
  F_1(k^2) &=& \dfrac{e^2 \Gamma(1-\epsilon_e)}{(4\pi)^{\epsilon_e+\frac{d_e}{2}}} 
 \left( \dfrac{1}{2(\de-2+\epe)(\de-3+\epe)} \right) \Bigg[ 2(\de-2)(3-\de-\epe)^2 \sint{1,0,\bepe}{\de} \nonumber \\
&&  + \Big( 4m^2[ 4-2\de +(2-4\de + \de^2)\epe + 2(\de-1) \epe^2 ] + k^2 [ \de^3 + 3\de^2(\epe-3) \nonumber \\
&& -8(\epe-2)^2 + 2\de(15-10\epe + \epe^2) ] \Big) \sint{1,1,\bepe}{\de} \Bigg]. \label{F1_2}
\end{eqnarray}
If we set~$\epe=0$ and use the identities in Eqs.~\eqref{sint2}, we arrive at the Dirac form factor for standard QED:
\begin{eqnarray}
F_1(k^2) &=& - \dfrac{e^2 }{m^2(4\pi)^{\frac{d_e}{2}}(\de-4)(\de-3)(k^2-4m^2)} \Bigg[ -2(\de-2)[k^2+(\de-5)(\de-2)m^2]\sint{0,1,0}{\de} \nonumber \\
&& -m^2(\de-3)[(16+\de^2-7\de)k^2-8m^2]\sint{1,1,0}{\de} \Bigg], \label{F1_SQED}
\end{eqnarray} 
which agrees with the result computed in the seminal work of Davydychev et. al. in Ref.~\cite{DavydychevVertex}. 
\end{widetext}

\subsection{The Pauli form factor} \label{Pauli_FormFactor}

For the simplification and eventual evaluation of the Pauli form factor, it is easy to appreciate that the combination of scalar integrals in Eq.~\eqref{F2} can be rewritten in a much simpler manner by a direct application of Feynman parameterization without the need to use elaborate expressions given in Appendix~\ref{scalar_integrals} to express it in terms of master integrals. Therefore, with some straightforward algebra, it can be shown that,
\begin{eqnarray}
&& \hspace{-15mm} F_2(k^2) = \dfrac{4 e^2 m^2 \Gamma\left( 3 - \epsilon_e - \frac{d_e}{2} \right)}{(4\pi)^{\epsilon_e+\frac{d_e}{2}}}  \nonumber \\
&& \hspace{-4mm} \times \int_0^1 dx \int_0^{1-x} dy\, \dfrac{2x+(2-d_e)(x^2+xy)}{(1-x-y)^{\epsilon_e} A^{3-\epsilon_e-\frac{d_e}{2}}} \,, \label{PF_FP}
\end{eqnarray}
where $A$ has been defined as
\begin{eqnarray}
A \equiv (x+y)^2 m^2 - xy k^2 +(1-x-y)m^2_\gamma \,. \label{defD}
\end{eqnarray}
Here we have introduced a photon mass regulator $m_\gamma$ to regularize infrared divergences. For $\epsilon_e=0$, which implies $\dg=\de$, the Pauli form factor in Eq.~\eqref{PF_FP} reads as,
\begin{eqnarray}
&& \hspace{-15mm} F_2^{S}(k^2) = \dfrac{4 e^2 m^2 \Gamma\left( 3 - \frac{d_e}{2} \right)}{(4\pi)^{\frac{d_e}{2}}} \nonumber \\
&& \hspace{-4mm} \times \int_0^1 dx \int_0^{1-x} dy\, \dfrac{2x+(2-d_e)(x^2+xy)}{ A^{3-\frac{d_e}{2}}} \,. \label{PF_std_QED}
\end{eqnarray}
It yields the standard (as the superscript ``S" suggests) result for $d_e=4$ and $d_e=3$ corresponding to conventional QED4 and QED3, respectively, (see for example Ref.~\cite{Guzman:2023hzq}). If we use the notation $F_2^{S-4}(k^2)$ for the Pauli form factor in 4 space-time dimensions for standard QED, we easily reproduce the result obtained first by Schwinger: 
\begin{eqnarray}
    F_2^{S-4}(0) &=& \dfrac{\alpha}
{2 \pi} \,.
\end{eqnarray}
Now notice that $d_\gamma=4$ and $d_e=3$ implies $\epsilon_e={1}/{2}$. In this case of RQED (as the notation below with the superscript $R$ signifies), the Pauli form factor of Eq.~\eqref{PF_FP} simplifies to the following integral form over the Feynman parameters:
\begin{eqnarray}
 && \hspace{-13mm}   F_2^R(k^2) = \dfrac{e^2}{4\pi^2}\int_0^1 dx\, \int_0^{1-x}dy\, \dfrac{2x-(x^2+x y)}{\sqrt{1-x-y}} \nonumber \\ \nonumber \\
    && \times \left( \dfrac{1}{(x+y)^2-x y c} \right) \,,
\end{eqnarray}
where  $c={k^2}/{m^2}$. Moreover, we have discarded the redundant photon mass regulator in this case as the result is convergent anyway.

After a convenient change of variables $y \xrightarrow{} x \left( {1}/{y} -1 \right) $ in the second integral over the $y$-variable, and interchanging the order of integration, we can readily integrate over the variable $x$. The above equation then reduces to:
\begin{eqnarray}
    F_2^R(k^2) = \dfrac{8 \alpha}{3\pi} \int_0^1 dy\, \dfrac{y}{1-c\,y(1-y)} \, , \label{F2R_1}
\end{eqnarray} 
where, as usual, the fine structure constant or the QED coupling is defined as $\alpha=e^2/(4\pi)$. Now making use of the simple identity 
\begin{eqnarray}
\int_0^1 dy\, \dfrac{y}{1 + c y (1-y)} = \dfrac{1}{2} \int_0^1 dy\, \dfrac{1}{1 + c y (1-y)} \,,
\end{eqnarray}
Eq.~\eqref{F2R_1} for the Pauli form factor for the case of RQED in $\dg=4$ and $\de=3$ dimensions for the photons and charged fermions, respectively, can be cast in the following equivalent form
\begin{eqnarray}
    F_2^R(k^2) = \dfrac{4 \alpha}{3\pi} \int_0^1 dy\, \dfrac{1}{1-c\,y(1-y)} \,. \label{F2R_2}
\end{eqnarray} 
Notice that $F_2^{S-4}(k^2)$ for the Pauli form factor in $4$ space-time dimensions for standard QED, Eq.~\eqref{PF_std_QED},  is trivially related to $F_2^R(k^2)$, Eq.~\eqref{F2R_2}, through the following multiplication  constant:
\begin{eqnarray}
    F_2^R(k^2) &=& \dfrac{8}{3} \, F_2^{S-4}(k^2) \,, \label{F2R_final}
\end{eqnarray}
since, \cite{Peskin},
\begin{eqnarray}
    F_2^{S-4}(k^2) = \dfrac{\alpha}{2\pi}  \int_0^1 \dfrac{dy}{1+cy(1-y)} \,.
\end{eqnarray}
Thus the Pauli form factor for the reduced theory is the same as in the standard QED in four dimensions scaled up by a factor of $\frac{8}{3}$.

We can now take the photon on-shell and consider the case $k^2=0$ which corresponds to the conventional definition of the anomalous magnetic moment of the fermion. Therefore, 
Eq.~\eqref{F2R_2} implies that,
\begin{eqnarray}
    F_2^R(0) = \dfrac{4\alpha}
{3\pi} \, . 
\end{eqnarray}
It agrees with the result reported in Ref.~\cite{RQED_g2} modulus the sign. As we reproduce known results in the literature for other cases and do not expect a change of sign in the value of the anomalous magnetic moment of the fermion as compared to the one for standard QED, we are confident of our result. 

\begin{figure}[h!]
\begin{center}

\begin{tikzpicture}[baseline=(current bounding box.north)]

\begin{feynman}

\vertex (n1);
\vertex [right=of n1] (n2);
\vertex [right=of n2] (n3);
\vertex [right=of n3] (n4);

\diagram*{
(n1) --[fermion, edge label=$p$] (n2) --[fermion] (n3) --[fermion,edge label=$p$] (n4),
(n3) --[charged boson, half right, edge label=$l$] n2,
}; 
\end{feynman}
\end{tikzpicture} 
\end{center}
\caption{One loop Feynman diagram for the fermion self energy $\Sigma(p)$.} \label{F_self}
\end{figure}
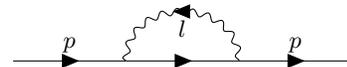

\begin{widetext}
\section{Fermion and photon self energies at one loop}\label{fermion_selfEnergy}

For the sake of completeness, we also evaluate one loop fermion self energy and the photon vacuum polarization.  Fig.~\ref{F_self} depicts the Feynman diagram for one loop fermion self energy function $\Sigma(p)$. According to the Feynman rules given in Fig.~\ref{Feynman_rules_RQED}, it reads as
\begin{eqnarray}
    -i \Sigma(p) = \dfrac{\Gamma(\bepe)}{(4\pi)^{\epe}} \int \dfrac{d^{\de}l}{(2\pi)^{\de}} \dfrac{i}{(-l^2)^{\bepe}} (-ie \gamma^{\mu}) \left[ \dfrac{i(\slashed{p}+\slashed{l}+m)}{(p+l)^2-m^2} \right] (-ie \gamma^\nu)  \left( \eta_{\mu\nu} - (1-\tilde{\xi}) \dfrac{l_\mu l_\nu}{l^2} \right),
\end{eqnarray}
which can be reorganized as follows:
\begin{eqnarray}
    -i \Sigma(p) = \dfrac{e^2\Gamma(\bepe)}{(4\pi)^{\epe}} \hspace{-2mm}  \int \hspace{-2mm} 
 \dfrac{d^{\de}l}{(2\pi)^{\de}} \dfrac{1}{[(p+l)^2-m^2](-l^2)^{\bepe}} \Bigg[ m(\de-1+\tilde{\xi}) -(\de-3+\tilde{\xi}) \slashed{p} 
     - \hspace{-1mm}  \left( \de-1-\tilde{\xi} + \dfrac{2(1-\tilde{\xi}) l\cdot p}{l^2} \hspace{-1mm}  \right) \slashed{l} \Bigg]. \label{fermion_se_l}
\end{eqnarray}
We can now implement the tensor reduction algorithm described in Refs.~\cite{Davydychev:1991va,Fiorentin,Anastasiou3} to express the fermion self energy in Eq.~\eqref{fermion_se_l} in terms of scalar integrals:
\begin{eqnarray}
    \Sigma(p) &=&  - \dfrac{e^2 \Gamma(\bepe)}{(4\pi)^{\epe + \frac{\de}{2}}} \Big\{ -m(\de-1+\tilde{\xi}) \sint{1,\bepe}{\de} + \big[(\de-3+\tilde{\xi})\sint{1,\bepe}{\de} -(\de-1-\tilde{\xi})\sint{2,\bepe}{\de+2} 
    \nonumber \\
    && \hspace{1cm} + (1-\tilde{\xi}) \sint{1,1+\bepe}{\de+2} - 4p^2(1-\tilde{\xi}) \sint{3,1+\bepe}{\de+4} \big]\slashed{p} \Big\} \,, \label{fermion_self_RQED}
\end{eqnarray}
\end{widetext}
where the scalar integrals $\sint{a,b}{D}$ with two labels are defined as follows
\begin{eqnarray}
    \sint{a,b}{D} = \int \dfrac{d^Dl}{i \pi^{D/2}} \dfrac{1}{[-(p+l)^2+m^2]^a(-l^2)^b} \,,
\end{eqnarray}
and where we have made use of the following well-known identities:
\begin{eqnarray}
    \int \dfrac{d^{\de}l}{i \pi^{\de/2}} \dfrac{l^\mu}{[(p+l)^2-m^2](-l^2)^b} &=& p^\mu \sint{2,b}{\de+2} \,, \nonumber \\
    \int \dfrac{d^{\de}l}{i \pi^{\de/2}} \dfrac{l^\mu l^\nu}{[(p+l)^2-m^2](-l^2)^b} &=& \dfrac{1}{2}\eta^{\mu\nu} \sint{1,b}{\de+2} \nonumber \\
    &-& 2 p^\mu p^\nu \sint{3,b}{\de+4} \,. 
\end{eqnarray}
Note that in the massless case, after performing the Feynman integrals, the fermion self energy of Eq.~\eqref{fermion_self_RQED}  reduces to a compact expression
\begin{eqnarray}
   && \hspace{-10mm} \Sigma(p) =  -\dfrac{e^2 \Gamma(\bepe)}{(4\pi)^{\epe + \frac{\de}{2}}} \dfrac{(\de-2)(-p^2)^{\frac{\de}{2}+\epe-2}}{2 \bepe (\de-2+\epe)} \times  \nonumber \\
    && \hspace{-6mm} \Big[ \bepe(\de-2)-(1-\tilde{\xi})(\de-2+\epe)\Big] G(\de,1,\bepe) \slashed{p} \,, \label{fermion_self_RQED_massless}
\end{eqnarray}
where the function $G(D,a,b)$ is defined according to
\begin{eqnarray}
  && \hspace{-8mm}  \sint{a,b}{D} = \int \dfrac{d^Dl}{i \pi^{\frac{D}{2}}} \dfrac{1}{[-(p+l)^2]^a (-l^2)^b} \nonumber \\
  && \hspace{-1mm} = (-p^2)^{\frac{D}{2}-a-b}\, G(D,a,b) \,, \\ ~ \nonumber \\
  && \hspace{-17mm}   G(D,a,b) = \dfrac{\Gamma\left(a+b-\frac{D}{2}\right) \Gamma\left( \frac{D}{2}-a \right)  \Gamma\left( \frac{D}{2}-b \right)}{\Gamma(a) \Gamma(b) {\Gamma(D-a-b)}}  \,.
\end{eqnarray}
Introducing the new variable:
\begin{eqnarray}
    \epsilon_\gamma &=& 2 - \dfrac{\de}{2}-\epe \,, \label{ep_gamma}
\end{eqnarray}
we can rewrite the expression in Eq.~\eqref{fermion_self_RQED_massless} for the massless self energy as
\begin{eqnarray}
    \Sigma(p) &=& - \dfrac{e^2 \Gamma(\bepe)(-p^2)^{\epsilon_\gamma}}{(4\pi)^{d_\gamma/2}} \Big[ \dfrac{2(1-\epsilon_\gamma-\epe)}{2-2\epsilon_\gamma-\epe} \nonumber \\
    &&- \xi (1-\epsilon_\gamma-\epe) \Big] G(\de,1,\bepe)\, \slashed{p} \,,
    \label{self_energy_massless}
\end{eqnarray}
which agrees with the earlier result presented in the literature, see Refs.~\cite{Teber2012,Teber_2018}.

In the case of standard QED with massive fermions, and which corresponds to $\epe = 0$ as stated earlier, the fermion self energy given in Eq.~\eqref{fermion_self_RQED} can be rewritten as follows:
\begin{eqnarray}
    \Sigma(p) &=& -\dfrac{e^2}{(4\pi)^{\frac{\de}{2}}} \Bigg\{m(1-\xi-\de) \sint{1,1}{\de}  \nonumber \\
    && + \dfrac{\de-2}{2p^2} \, \xi  \, \Big[ \sint{1,0}{\de} + (m^2+p^2) \sint{1,1}{\de} \Big]  \Bigg\} \,, \label{self_energy_std_QED}
\end{eqnarray}
where we have used the identities given in the equations \eqref{two_point_stan_QED} in the corresponding appendix. 
This resulting expression of 
Eq.~\eqref{self_energy_std_QED} for the massive fermion self energy in standard QED agrees with the known result in the literature (see for example \cite{DavydychevVertex,Guzman:2023hzq, WL_fprop1}). 


\begin{figure}[h!]
\begin{center}

\begin{tikzpicture}[baseline=(current bounding box.north)]

\begin{feynman}

\vertex (n1);
\vertex [right=of n1] (n2);
\vertex [right=of n2] (n3);
\vertex [right=of n3] (n4);

\diagram*{
(n1) --[charged boson, edge label=$p$] (n2) --[fermion, half left, edge label=$p+l$] (n3) --[charged boson,edge label=$p$] (n4),
(n3) -- [fermion, half left, edge label=$l$] (n2)
}; 
\end{feynman}
\end{tikzpicture} 
\end{center}
\caption{One loop Feynman diagram for the photon self energy function $\Pi(p)^{\mu\nu}$.} \label{P_self}
\end{figure}
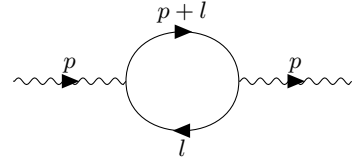
We now proceed with the calculation of the photon self energy tensor $\Pi^{\mu\nu}(p)$ at one loop. Fig.~\ref{P_self} depicts the Feynman diagram for this function. According to the Feynman rules in Fig.~\ref{Feynman_rules_RQED}, the function $\Pi^{\mu\nu}(p)$ reads as
\begin{eqnarray}
    i \Pi^{\mu\nu}(p) &=& (-1) (-ie)^2 \int \dfrac{d^{\de}l}{(2\pi)^{\de}} \times  \nonumber \\
    && \text{Tr} \Bigg[ \gamma^\mu \dfrac{i(\slashed{l}+m)}{l^2-m^2} \gamma^{\nu} \dfrac{i(\slashed{p}+\slashed{l}+m)}{(l+p)^2-m^2} \Bigg] \,.
\end{eqnarray}
Defining the function $\Pi(p^2)$ as
\begin{eqnarray}
    \Pi(p^2) = \dfrac{\Pi^{\mu}_{\mu}}{p^2(\de-1)} \,,
\end{eqnarray}
and carrying out the standard calculations, we obtain the result
\begin{eqnarray}
    \Pi(p^2) = -\dfrac{2 \, d_e \, e^2 \Gamma\left(2-\frac{\de}{2}\right)}{(4\pi)^{\frac{\de}{2}}} \int_0^1 dx\, \dfrac{x(1-x)}{\Delta^{2-\frac{\de}{2}}} \,, \label{photon_self_energy}
\end{eqnarray}
with 
\begin{eqnarray}
    \Delta = m^2-x(1-x)p^2 \,,
\end{eqnarray}
 where we have considered the gamma matrices $\gamma^\mu$ to have dimensions $d \times d$. The expression in Eq.~\eqref{photon_self_energy} agrees with the textbook result when we identify the dimensions of the gamma matrices with the space-time dimensions under consideration (see for example~\cite{Peskin}). 


\section{Renormalization constants } \label{RenormalizationConstants}

When $\dg=\de=4$, and $\dg=4$ with $\de=3$, the scalar integrals $\sint{1,\bepe}{\de}$, $\sint{2,\bepe}{\de+2}$ and $\sint{1,1+\bepe}{\de+2}$ in equation \eqref{fermion_self_RQED} for the fermion self energy are ultraviolet (UV) divergent. For the case of the vertex function $V^\mu$ the only UV divergent scalar integrals in Eqs.~\eqref{VL} and~\eqref{VT} are $\sint{1,0,\bepe}{\de}$ and $\sint{1,1,\bepe}{\de+2}$. Note that the divergences of the fields, charge and mass can be absorbed into the renormalization constants ${\cal{Z}}_1$, ${\cal{Z}}_2$, ${\cal{Z}}_3$ and ${\cal Z}_m$ which relate the bare fermionic field $\psi$, the bare coupling constant $e$, the bare photon field $A^\mu$ and the fermion mass $m$, respectively, to the renormalized fermion field $\psi_r$, charge $e_r$, photon field $A_r^\mu$ and mass $m_r$ as:
\begin{align}
\psi &= {\cal{Z}}_2 \psi_r \,, &\quad e &= {\cal{Z}}_1 e_r \,, \nonumber \\
A^{\mu} &=  {\cal{Z}}_3 A^{\mu}_r \,, &\quad m &= {\cal Z}_m m_r \,. 
\end{align}
Let us proceed to evaluate these constants. Notice that after the Feynman parameterization, the 
two-point scalar integrals $\sint{a,b}{D}(p)$  can be rewritten as:
\begin{eqnarray}
    \sint{a,b}{D}(p) = \dfrac{\Gamma\left( a+b-\frac{D}{2} \right)}{\Gamma(a) \Gamma(b)} \int_0^1 dx\, \dfrac{x^{a-1}(1-x)^{b-1}}{B^{a+b-\frac{D}{2}}} \,, 
\end{eqnarray}
where 
\begin{eqnarray}
    B = x(x-1)p^2+x\,m^2+y\,m_\gamma^2 \,. 
\end{eqnarray}
Thus the ultraviolet (UV) divergent parts of the fermion self energy in Eq.~\eqref{fermion_self_RQED} read as
\begin{eqnarray}
&& \hspace{-7mm}   \Sigma_{\text{UV}}^{S-4} = - \dfrac{e^2}{(4\pi)^2} \left[ -m\left( \dfrac{3+\xi}{\epsilon_\gamma} \right) + \dfrac{\xi}{\epsilon_\gamma} \slashed{p} \right], \nonumber \\
&& \hspace{-7mm}     \Sigma_{\text{UV}}^{R} = - \dfrac{e^2}{(4\pi)^2}  \left[ -m\left( \dfrac{4+2\tilde{\xi}}{\epsilon_\gamma}\right) + \left( \dfrac{6\tilde{\xi}-4}{3 \epsilon_\gamma} \right)\slashed{p} \right]. \label{fermion_self_UV}
\end{eqnarray}
Here $\Sigma_{\text{UV}}^{S-4}$ represents the UV divergent part for standard QED where we have fixed $\epe=0$, and $\de = 4-2\epsilon_\gamma$ while $\Sigma_{\text{UV}}^{R}$ represents the UV divergent part for RQED were we have set $\epe=\frac{1}{2}$, and $\de = 3-2\epsilon_\gamma$. 


Therefore, one can readily compute the renormalizations constants ${\cal{Z}}_2$ from Eqs.~\eqref{fermion_self_UV} at the one loop order in the $\overline{\text{MS}}$ scheme:
\begin{eqnarray}
    {\cal Z}_2^{S-4} &=& 1 - \dfrac{\alpha }{4\pi} \left( \dfrac{\xi}{\epsilon_\gamma} \right) \,, \nonumber \\
    {\cal Z}_2^{R} &=& 1 - \dfrac{\alpha}{6\pi}  \left( \dfrac{3\tilde{\xi}-2}{\epsilon_\gamma} \right) \,, \label{Z2_constants}
\end{eqnarray}
which agree, respectively,  with the well known result for standard QED in $\de=4$ dimension, see for example \cite{Srednicki}, and for the massless result in RQED in $\de=3$ and $\dg=4$ dimensions reported in Ref.~\cite{Teber2012}.

Similarly, drawing on Eqs.~\eqref{fermion_self_UV} again, we obtain the following fermion mass renormalizations constants in the $\overline{\text{MS}}$ scheme:
\begin{eqnarray}
    {\cal Z}_m^{S-4} &=& 1 - \dfrac{\alpha}{4\pi}\left(\dfrac{3+\xi}{\epsilon_\gamma} \right), \nonumber \\
    {\cal Z}_m^{R} &=& 1 - \dfrac{\alpha}{4\pi}\left( \dfrac{4+2\tilde{\xi}}{\epsilon_\gamma} \right).
\end{eqnarray}
Now, since $\sint{1,0,\bepe}{\de}$ and $\sint{1,1,\bepe}{\de+2}$ are the only UV divergent scalar integrals that show up in the vertex, the term $(\de-3+\tilde{\xi})\sint{1,0,\bepe}{\de}+(2-\de)\sint{1,1,\bepe}{\de+2}$ in equation \eqref{VL} is the unique contribution to the UV-divergent part for the vertex $V^\mu$. Thus, following a similar analysis as presented for the fermion self energy, the UV-divergent parts for $V_\mu$, within the dimensional regularization scheme, read as
\begin{eqnarray}
    V_{\mu}^{S-4} &=& \dfrac{e^3}{(4\pi)^2} \dfrac{\xi}{\epsilon_\gamma}\, \gamma_\mu \,, \nonumber \\
    V_\mu^{R} &=& \dfrac{e^3}{(4\pi)^2} \left( \dfrac{6\tilde{\xi} - {4}}{3 \epsilon_\gamma} \right) \, \gamma_\mu \,. \label{UV_vertex}
\end{eqnarray}
The first equation corresponds to the standard QED in 4 dimensions, and the second equation to RQED in $\dg=4$ and $\de=3$.
We can identify ${\cal{Z}}_1$ through Eq.~\eqref{UV_vertex}, concluding that at the one loop order for both the standard QED as well as RQED:
\begin{eqnarray}
    {\cal Z}_1 = {\cal Z}_2 \,,
\end{eqnarray}
in agreement with the non-perturbative Ward-Takashi identity. 

Finally, according to Eq.~\eqref{photon_self_energy} the computation of ${\cal{Z}}_3$ at one loop proceeds as in standard QED in $\de = 4$ dimensions, while for $\de = 3$ it vanishes at one loop, since the photon self energy is finite at that order. Thus, since ${\cal Z}_1 = {\cal Z}_2$, the beta function $\beta(\alpha)$ receives 
no corrections at this order of approximation which agrees with the result reported in Ref.~\cite{Teber2012}. 

\section{Summary and conclusions}\label{conclusions}

In this article, we obtain general expressions for the vertex function $V^\mu(p',p)$ in general and Dirac and Pauli form factors $F_1(k^2)$ and $F_2(k^2)$, respectively, in particular
 in arbitrary covariant gauges and dimensions at one loop order in perturbation theory for QED.
We allow the bulk gauge field to reside in $\dg$ space-time dimensions interacting with a fermionic field whose dynamics might be restricted to different 
space-time dimensions $\de$, with $\de \leq \dg$. The main feature of this work is the novel computational tool available to us by a combination of first and second order formulation of QED as was employed in our previous work~\cite{Guzman:2023hzq}.
As stated earlier, this coupled formalism has numerous advantageous as compared to the usual solely first order formulation of QED calculations:
\begin{enumerate}
    \item It renders the calculation efficient.
    \item It  implements Ward-Takahashi identity and decomposes the fermion-photon vertex naturally into its longitudinal and transverse components with respect to the photon momentum without resorting to the customary Ball-Chiu construction.  
    \item It allows us to systematically demonstrate that the Pauli form factor for on-shell fermions is independent of the gauge parameter for this general setup of potentially different $\dg$ and $\de$ dimensions by invoking identities which stem from the combined formalism.   
    \item The extraction of the anomalous magnetic moment of the charged fermions is greatly expedited and we hope this combined formalism will be of practical usage in the computation of this and other physical observables at higher loops.  
\end{enumerate}

For the sake of completeness, we also compute fermion self energy $\Sigma$ and photon vacuum polarization tensor $\Pi^{\mu\nu}$, respectively.  
The expressions for the longitudinal and transverse parts of the vertex function $V^\mu(p',p)$, the Dirac form factor $F_1(k^2)$, the Pauli form factor $F_2(k^2)$, the fermion self energy $\Sigma(p)$ and the photon vacuum polarization tensor $\Pi^{\mu\nu}$ can be referred to in Eqs.~\eqref{VL},~\eqref{VT},~\eqref{F1_2},~\eqref{PF_FP},~\eqref{fermion_self_RQED} and ~\eqref{photon_self_energy}, respectively, in terms of Feynman scalar integrals. Starting from these equations, as has been detailed all along the article, one can easily reproduce  known results for standard QED where $\dg=\de$, and for massless RQED. Though some of the connections are rather trivial, it might be worth summarizing our cross-checks and collecting them in the following list:
\begin{itemize}
    \item Setting $\epe = 0$ in Eqs.~\eqref{V_1L} and~\eqref{VT}, we obtain the same expressions found in Ref.~\cite{Guzman:2023hzq} for the longitudinal and transverse components of the vertex function $V^\mu$ in standard QED.    
    \item The Dirac form factor for standard QED, Eq.~\eqref{F1_SQED}, agrees with the results reported in~\cite{DavydychevVertex}.
    \item The Pauli form factor for standard QED is displayed in Eq.~\eqref{PF_std_QED} and it agrees with the well-known results in the literature (see for example \cite{Guzman:2023hzq}).
    \item The fermion self energy for standard QED, i.e., Eq.~\eqref{self_energy_std_QED}, is in agreement with the known result in the literature (see for example \cite{DavydychevVertex,Guzman:2023hzq, WL_fprop1}). 
    \item The fermion self energy for massless RQED, i.e.,  Eq.~\eqref{self_energy_massless}, agrees with the result presented in the Refs.~\cite{Teber2012,Teber_2018}.
    \item The computation of the photon self energy function at one loop yields the same result as in the conventional QED, Eq.~\eqref{photon_self_energy}.
\end{itemize}

From the results obtained in the article, we can compare and contrast the difference between standard QED and RQED. For example, as shown in section~\ref{Pauli_FormFactor}, the Pauli form factor for the massive RQED, $\dg=4$ and $\de=3$, is $8/3$ times the form factor of standard QED in four dimensions (see Eq.~\eqref{F2R_final}). One may be tempted to attribute it to the observation that the response of the gyromagnetic ratio $g$ of the charged fermion to a magnetic field in RQED does not get diluted into the third space dimension and gets augmented.    

Additionally, from our general expressions in Eqs.~\eqref{VL}, \eqref{VT} and~\eqref{fermion_self_RQED}, we can easily identify the ultraviolet divergences in the QED theories, observing that the only scalar integrals with ultraviolet divergences when $\dg=\de=4$, and $\dg=4$, $\de=3$, are $\sint{1,0,\bepe}{\de}$, $\sint{1,1,\bepe}{\de+2}$, $\sint{1,\bepe}{\de}$, $\sint{2,\bepe}{\de+2}$ and $\sint{1,1+\bepe}{\de+2}$. Extracting the singular parts of these integrals in dimensional regularization, we obtain the result displayed in Eq.~\eqref{Z2_constants} for the renormalization constant ${\cal{Z}}_2$. We also show that ${\cal{Z}}_1 = {\cal{Z}}_2$ at one loop, as expected by the Ward-Takahashi identity which relates the fermion propagator with the fermion-photon vertex. This result, together with the observation that the photon self energy in $\de=3$ dimensions is finite, Eq.~\eqref{photon_self_energy}, implies that the beta function for RQED in $\dg=4$ and $\de=3$ vanishes at one loop order. 

Moreover, our computation of the three-point fermion-photon interaction vertex at one loop order in RQED in arbitrary gauges and dimensions may provide a natural guide for any non-perturbative construction of this Green function which is a highly sought after goal in the Schwinger-Dyson equation studies of non-perturbative field theories, see for example~\cite{Curtis:1993py,Burden:1993gy,Bashir:1994az,Bashir:2000rv,Bashir:2002dz,Bashir:2004hh,Bashir:2004yt,Bashir:2007qq,Kizilersu:2009kg,Bashir:2011vg,Bashir:2011dp,Kizilersu:2014ela,Albino:2018ncl,Albino:2021rvj,Albino:2022efn,Lessa:2022wqc}. All this is for future. 

\begin{acknowledgments}
V.M.B.G. is grateful to {\em Consejo Nacional de Ciencia y Tecnolog\'ia (M\'exico)} for support. A.B. acknowledges {\em Coordinaci\'on de la Investigaci\'on Cient\'ifica} of the {\em Universidad Michoacana de San
Nicol\'as de Hidalgo} grant 4.10 and the Fulbright-Garc\'ia
Robles scholarship for his stay as a visiting scholar at the
Thomas Jefferson National Accelerator Facility, Newport
News, Virginia, USA. 
\end{acknowledgments}


\appendix

\section{Conventions} \label{conventions}

Our Minkowski space metric $\eta^{\mu\nu} = \text{diag}(+,-,-,-)$, while the Dirac gamma matrices are in general $d \times d$ matrices, where $d$ equals to the space-time dimension $\de$ of the fermion when $\de$ is even, and different when $\de$ is odd.  The Dirac gamma matrices satisfy the anticommutation relation $\{\gamma^\mu, \gamma^\nu \} = 2 \eta^{\mu\nu}$, and $\text{Tr}(\gamma^\mu\gamma^\nu) = d \,\eta^{\mu\nu}$.

\section{Three-point on-shell scalar 
 integrals} \label{scalar_integrals}

Using the dimensional recurrence relations for $n$-point one loop integrals discussed in Ref. \cite{IntDimRed}, we can obtain the following on-shell integral identity which transforms scalar integrals in $D+2$ dimensions into scalar integrals in $D$ dimensions,
\begin{eqnarray}
J_{a,b,c}^{D+2}(p,p^\prime) = \dfrac{J^D_{a,b,c-1}(p,p^\prime)}{(D-a-b-c+1)}, \label{dim_rec_iden}
\end{eqnarray}
where $J_{a,b,c}^{D}$ is defined in Eq.~\eqref{scalar_int}.

We can obtain further relations between on-shell scalar integrals through the implementation of the now widely used integration by parts technique (IBP), \cite{IBP1,IBP2}. This method yields the following useful identities,
\begin{eqnarray}
&& \hspace{3mm} \sint{a,b,c}{\de} = \dfrac{a\sint{a+1,b,c-1}{\de} + b \sint{a,b+a,c-1}{\de} }{\de-a-b-2c} \,, \label{IBP1} \\ \nonumber \\
&& \hspace{-11.6mm} 2a p\cdot p^\prime\,\sint{a+1,b,c}{\de}  = a \left( \sint{a+1,b-1,c}{\de} -  \sint{a+1,b,c-1}{\de} \right)  \nonumber \\
&& \hspace{12mm}  + \, b \left( \sint{a,b-1,c+1}{\de} -  \sint{a,b+1,c-1}{\de} \right) \nonumber \\ 
&& \hspace{12mm}   - \, (c-b) \sint{a,b,c}{\de} -  2bm^2\,\sint{a,b+1,c}{\de} \, , 
\label{IBP2}
\end{eqnarray}  
where we have suppressed the $p$ and $p^\prime$ dependence for notational simplicity.

By combining the above identities \eqref{IBP1} and \eqref{IBP2}, we arrive at the following practically useful relation. Note that it leaves the third label in the on-shell scalar integrals unchanged:
\begin{eqnarray}
&& \hspace{-10mm} \sint{a,b,c}{\de} = \dfrac{1}{\beta} \Big[ \alpha_1 \sint{a-1,b,c}{\de} - \alpha_2 \sint{a-1,b+1,c}{\de} + \alpha_3 \sint{a,b-1,c}{\de} \Big] \,,  \label{IBP3}  
\end{eqnarray}
where 
\begin{eqnarray}
    \alpha_1 &=& (1+\de-a-2b-2c) (\de-a-b-c) \,, \nonumber \\ 
    \alpha_2 &=& 2m^2b(a+b+2c-\de) \,, \nonumber \\
    \alpha_3 &=& (a-1)(a+b+2c-\de) \,, \nonumber \\ 
    \beta &=& 2 p \cdot p^\prime (a-1)(a+b+2c-\de) \,.
\end{eqnarray}
Using the identities in Eqs.~\eqref{dim_rec_iden}-\eqref{IBP3}, the scalar integrals that appear in the Dirac and Pauli form factors acquire the following form:
\begin{eqnarray}
\sint{1,1,1+\bepe}{d_e+2} &=& \dfrac{1}{d_e-3+\epsilon_e}\, \sint{1,1,\bepe}{d_e} \,, \nonumber  \\ \nonumber  \\ 
\sint{2,1,\bepe}{\de+2} &=& \dfrac{\de-2(1+\bepe)}{2(\de-3+\epe)}\, \sint{1,1,\bepe}{\de} \,, \nonumber 
\end{eqnarray}
\begin{eqnarray}
\sint{2,2,1+\bepe}{\de+4} &=& \dfrac{1}{4(m^4-(p \cdot p^\prime)^2)(\de-3+\epe)} \Big[ \Big( p\cdot p^\prime(\de \nonumber \\
&&  -6  +2\epe)+m^2(4-\de-2\epe) \Big) \sint{1,1,\bepe}{\de}  \nonumber \\
&& + 2m^2 \sint{0,2,\bepe}{\de} \Big] \,, \nonumber \\
\sint{3,1,1+\bepe}{\de+4} &=& \dfrac{1}{8(m^4-(p \cdot p^\prime)^2)(\de-3+\epe)} \Big[ \Big( p\cdot p^\prime(\de \nonumber \\
&& -4 +2\epe)+m^2(6-\de-2\epe) \Big) \sint{1,1,\bepe}{\de} \nonumber \\
&& - 2 p \cdot p^\prime \sint{0,2,\bepe}{\de} \,, \nonumber \\
\sint{1,1,\bepe}{\de+2} &=& \dfrac{1 }{(\de-3+\epe)(\de-2+\epe)} \Big[ (\de-3 +\epe) \nonumber \\
&\times&  \sint{1,0,\bepe}{\de} -(m^2+p \cdot p^\prime)(\de-4+2\epe)\sint{1,1,\bepe}{\de} \Big] \,, \nonumber \\ \nonumber \\
\sint{1,2,\bepe}{\de+2} &=& \dfrac{\de-4+2\epe}{2(\de-3+\epe)} \sint{1,1,\bepe}{\de} \,, \nonumber \\ \nonumber \\
\sint{2,2,\bepe}{\de+4}  &=& \dfrac{(\de-3+2\epe)(\de-4+2\epe)}{2(\de-3+\epe)(\de-2+\epe)} \sint{1,1,\bepe}{\de} \nonumber \\
&& -2 \sint{1,3,\bepe}{\de+4} \,. \label{sint1}
\end{eqnarray}
We can also derive the following useful relations for the standard QED case,
\begin{eqnarray}
\sint{0,2,1}{\de} &=& \dfrac{\de-2}{4m^4} \sint{0,1,0}{\de} \,, \nonumber  \\
\sint{1,0,1}{\de} &=& - \dfrac{\de-2}{2m^2(\de-3)} \sint{0,1,0}{\de} \,, \nonumber \\
\sint{1,1,1}{\de} &=& - \dfrac{\de-2}{2(\de-4)m^2(m^2+p \cdot p^\prime)} \sint{0,1,0}{\de} \nonumber \\
&& - \dfrac{\de-3}{(\de-4)(m^2+p \cdot p^\prime)} \sint{1,1,0}{\de} \,. \label{sint2}
\end{eqnarray}
We now focus our attention on the relatively simpler two-point scalar integrals in the next Appendix.

\section{Two-point scalar integrals} \label{scalar_integrals_2}

Making use of the dimensional recurrence relations for $n$-point one loop integrals discussed in Ref. \cite{IntDimRed}, we readily obtain
\begin{eqnarray}
    \sint{a,b}{D+2} &=& \dfrac{1}{2p^2(D-a-b+1)} \Big[ (m^2-p^2)^2 \sint{a,b}{D} \nonumber \\
    &-& (m^2-p^2) \sint{a-1,b}{D} + (m^2+p^2)\sint{a,b-1}{D} \Big] \,.  \label{int2_dimred}
\end{eqnarray}
For these types of Feynman integrals we obtain the following recursive relations from the IBP procedure, 
\begin{eqnarray}
    0 &=& (\de-a-2b)\sint{a,b}{D} - a \sint{a+1,b-1}{D} + a(m^2-p^2)\sint{a+1,b}{D} \,, \nonumber \\
    0 &=& -b \sint{a-1,b+1}{D} +(b-a) \sint{a,b}{D} + b(m^2-p^2)\sint{a,b+1}{D} \nonumber \\
    && + a \sint{a+1,b-1}{D} + a (m^2+p^2) \sint{a+1,b}{D} \,.
\end{eqnarray}
From these relations we can further derive the following identities:
\begin{eqnarray}
    \sint{a+1,b}{D} &=& \dfrac{1}{2am^2} \Big[ b \sint{a-1,b+1}{D} + b(p^2-m^2)\sint{a,b+1}{D} \nonumber \\
    && +(2a+b-D)\sint{a,b}{D}\Big] \,,  \nonumber \\
    \sint{1,b+1}{D} &=& \dfrac{1}{b(m^2-p^2)} \Big[ (1-b)\sint{1,b}{D} - \sint{2,b-1}{D} \nonumber \\
    && -(m^2+p^2) \sint{2,b}{D} \Big] \,.
\end{eqnarray}
Combining these equations with Eq.~\eqref{int2_dimred}, we can also obtain
\begin{eqnarray}
    \sint{2,1}{\de+2} &=& - \dfrac{1}{2p^2} \left( \sint{1,0}{\de} + (m^2-p^2) \sint{1,1}{\de} \right) \,, \nonumber \\ \nonumber \\
    \sint{1,2}{\de+2} &=& \dfrac{1}{2p^2} \left( \sint{1,0}{\de} + (m^2+p^2) \sint{1,1}{\de} \right) \,, \nonumber \\ 
    \sint{3,2}{\de+4} &=& \dfrac{1}{8p^4} \left( \de \sint{1,0}{\de} + [-4p^2+\de(m^2+p^2)]\sint{1,1}{\de} \right) \,. \nonumber \\ \label{two_point_stan_QED}
\end{eqnarray}

\section{LKF transformation for the fermion propagator in massive RQED}

The transformation of the fermion propagator in coordinate space under a variation of the gauge parameter is given by the LKF transformation, which for RQED reads 
as~\cite{Ahmad:2016dsb},
\begin{eqnarray}
    S_{F}(x,\xi) = S_{F}(0,\xi)\,e^{-i\left[ \tilde{\Delta}_{\de}(0,\epsilon_e) - \tilde{\Delta}_{\de}(x,\epsilon_e) \right]} \,, \label{LKF_RQED}
\end{eqnarray}
where $S_{F}(x,\xi)$ is the fermion propagator in coordinate space in an arbitrary covariant gauge $\xi$, and
\begin{eqnarray}
    \tilde{\Delta}_{\de}(x,\epsilon_e) = -i \dfrac{\bepe\Gamma[\bepe]}{(4\pi)^{\epe}} \xi e^2 \dfrac{ \Gamma\left( \dfrac{\de-a}{2} \right) }{2^a\pi^{\de/2}\Gamma\left(\dfrac{a}{2}\right)}\, \left(\mu x \right)^{a-\de} \,, \nonumber \\
\end{eqnarray}
where $a=4-2\epe$, and $\mu$ is a regulator with mass scale.

For the case of RQED in $\dg=4$ and $\de=3$ dimensions, the exponent in the exponential factor in the Eq.~\eqref{LKF_RQED} reduces to~\cite{Ahmad:2016dsb},
\begin{eqnarray}
    -i\left[ \tilde{\Delta}_{3}(0,\epsilon_e) - \tilde{\Delta}_{3}(x,\epsilon_e) \right] = \ln \left( \dfrac{x}{x_{\text{min}}} \right)^{-2\nu} \,, \label{Delta_LKF}
\end{eqnarray}
where $x_{\text{min}}$ is a regulating cutoff in the coordinate space, and $\nu=\alpha \xi  /(4\pi) $.

Now, the fermion propagator in momentum space can be written as follows in its most general form:
\begin{eqnarray}
    S_F(p;\xi) \equiv A(p;\xi) + B(p;\xi) \slashed{p} \equiv \dfrac{F(p;\xi)}{\slashed{p}-M(p;\xi)} \,, \label{F_prop_momentum}
\end{eqnarray}
where we have explicitly expressed the dependence on momentum $p$ and gauge parameter $\xi$. The scalar functions $A(p;\xi)$ and $B(p;\xi)$ are related to the wave function renormalization  $F(p;\xi)$ and the mass function $M(p;\xi)$  according to:
\begin{eqnarray}
    M(p;\xi) &=& \dfrac{A(p;\xi)}{B(p;\xi)} \,, \nonumber \\
    F(p;\xi) &=& B(p;\xi) p^2 - \dfrac{A(p;\xi)^2}{B(p;\xi)} \,. \label{F_M_functions}
\end{eqnarray}
Using Eqs.~\eqref{LKF_RQED} and~\eqref{Delta_LKF}, and following the procedure outlined in Ref.~\cite{Bashir:2002sp}, we can obtain non-perturbative expressions for the functions $A(p;\xi)$ and $B(p;\xi)$ defined through Eq.~\eqref{F_prop_momentum}. Thus, motivated by the lowest order perturbation theory, we take:
\begin{eqnarray}
    F(p;0) &=& 1 \,, \nonumber \\
    M(p;0) &=& m \,. \label{def_FM_p0}
\end{eqnarray}
Employing Eqs.~\eqref{LKF_RQED},~\eqref{Delta_LKF} and~\eqref{def_FM_p0}, we obtain
\begin{eqnarray}
    S_F(p;\xi) &=& \int d^3x \, e^{ipx}\Big[ \int \dfrac{d^3q}{(2\pi)^3} e^{-iqx} \left( \dfrac{\slashed{q}+m}{q^2-m^2} \right) \nonumber \\
    && \times \left(\dfrac{x_{\text{min}}}{x}\right)^{2\nu} \Big] \,.
\end{eqnarray}
Performing first the position space integration,
\begin{eqnarray}
    S_F(p;\xi) = C \int \dfrac{d^3q}{(2\pi)^3} \dfrac{\slashed{q}+m}{[(p-q)^2]^{\frac{3}{2}-\nu}(q^2-m^2)} \,, \label{SF_intq_LKF}
\end{eqnarray}
where
\begin{eqnarray}
    C = \dfrac{2^{3-2\nu}x_{\text{min}}\pi^{\frac{3}{2}}}{\Gamma(\nu)} \Gamma\left(\dfrac{3}{2}-\nu\right) \,.
\end{eqnarray}
After a direct application of Feynman parameterization, some standard calculations and using Eq.~\eqref{SF_intq_LKF}, we obtain the following expressions for the scalar functions $A^{\text{LKF}}(p;\xi)$ and $B^{\text{LKF}}(p;\xi)$ in terms of one Feynman integration parameter:
\begin{eqnarray}
    A^{\text{LKF}}(p;\xi) &=& c\, m \int_0^1 dy\, \dfrac{y^{\frac{1}{2}-\nu}}{D^{1-\nu}} \,, \nonumber \\
    B^{\text{LKF}}(p;\xi) &=& c\,\int_0^1 dy\, \dfrac{y^{\frac{3}{2}-\nu}}{D^{1-\nu}} \,, \label{AB_1Fp}
\end{eqnarray}
where
\begin{eqnarray}
    c &=& - \dfrac{e^{-i\pi\nu} x_{\text{min}}^{2\nu} \Gamma(1-\nu) }{2^{2\nu} \Gamma(\nu)} \,, \nonumber \\
    D &=& (1-y)(m^2-yp^2) \,.
\end{eqnarray}
Here, the superscript label LKF on the scalar functions $A(p;\xi)$ and $B(p;\xi)$ indicates that these functions are obtained from the LKF transformation.  

Making use of Eqs.~\eqref{F_M_functions} and~\eqref{AB_1Fp} and then performing integration on the variable $y$, we obtain the following non-perturbative expressions for $F(p;\xi)$ and $M(p;\xi)$ for RQED in $\dg=4$ and $\de=3$ dimensions,
\begin{widetext}
\begin{eqnarray}
    F^{\text{LKF}}(p;\xi) &=& - (p^2-m^2)^\nu\, \dfrac{x_{\text{min}}^{2\nu} \nu\,\Gamma(-2\nu)}{p(1+2\nu) G^\nu} \left[m^2+p^2 + (p^2-m^2)G^{2\nu} + 4\nu\,mp \right] \nonumber \\
    && \times \left[ \dfrac{p^2-m^2+(m^2+p^2-4\nu\, m\, p)G^{2\nu}}{m^3+p^3 + (p-m)(m^2+p^2+(1-2\nu)m\,p) G^{2\nu} +2\nu\, m\, p (m+p) } \right], \nonumber \\ ~ \nonumber \\
    M^{\text{LKF}}(p;\xi) &=& \dfrac{(1+2\nu)p\sqrt{G}\left(G^{2\nu-1}-1\right)(m^2-p^2)}{m\left(G^{2\nu}-1\right)(m^2+2\nu\,p^2)-p\left(G^{2\nu}+1\right)(p^2+2\nu\,m^2)} \left( 1 - \dfrac{p^2}{m^2} \right)^{-\frac{1}{2}}, \label{MF_RQED}
\end{eqnarray}    
\end{widetext}
where $p\equiv \sqrt{p^2}$, and the function $G$ is defined as
\begin{eqnarray}
    G = \dfrac{2m}{m-p}-1 \,.
\end{eqnarray}
Taking massless limit $m \rightarrow 0 $ in Eqs.~\eqref{MF_RQED}, we obtain the following expressions:
\begin{eqnarray}
    F^{\text{LKF}}(p;\xi) &=& \dfrac{(p\, x_{\text{min}})^{2\nu} \cos(\pi\nu) \Gamma(1-2\nu)}{1+2\nu} \,, \nonumber \\
    M^{\text{LKF}}(p;\xi) &=& 0 \,,
\end{eqnarray}
which agree with the results presented in Ref.~\cite{Ahmad:2016dsb}.
Since we want to compare with the one-loop results in the main text, we take a linear expansion in $\nu$ in Eqs.~\eqref{MF_RQED}. Thus
\begin{eqnarray}
    F^{\text{LKF}}(p;\xi) &=& 1 + \Big[ -2 + 2\gamma_E - \dfrac{2m^2}{p^2} + \dfrac{m^3 \ln(G)}{p^3} \nonumber \\
    && + \ln\left(\dfrac{p^2-m^2}{\Lambda^2}\right) + \ln(4) \Big]\nu + O(\nu^2) \,, \nonumber \\
    M^{\text{LKF}}(p;\xi) &=& m \left[ 1 + \dfrac{(m^2-p^2)\left( m \ln(G)-2p \right)}{p^{3}}\nu \right] \nonumber \\
    && + O(\nu^2) \,, \label{FM_LKF_LinearO}
\end{eqnarray}
where we have set $\Lambda = 2/x_{\text{min}}$, and $\gamma_E$ is the Euler's constant. Following perturbative approach, the fermion propagator can be written as,
\begin{eqnarray}
    S_F(p;\xi) = \dfrac{1}{\slashed{p}-m-\Sigma(p;\xi)} \,,
\end{eqnarray}
where $\Sigma(p;\xi)$ is the fermion self energy. Comparing the equation above with Eq.~\eqref{F_prop_momentum}, we obtain 
\begin{eqnarray}
    F(p;\xi) &=& 1 + \dfrac{1}{4p^2} \text{Tr}\left(\slashed{p} \Sigma_{1-\text{Loop}} \right) + O(\alpha^2) \,, \nonumber \\
    M(p;\xi) &=& m +\dfrac{m}{4p^2}\text{Tr}\left(\slashed{p} \Sigma_{1-\text{Loop}}\right) \nonumber \\ 
    && + \dfrac{1}{4}\text{Tr}\left( \Sigma_{1-\text{Loop}}\right) + O(\alpha^2) \,,  \label{FM_1Loop}
\end{eqnarray}
where $\Sigma_{1-\text{Loop}}$ is the one-loop result given in Eq.~\eqref{fermion_self_RQED}, with $\de=3-2\epsilon_\gamma$ and $\bepe=\frac{1}{2}$. To evaluate the scalar integrals involved in $\Sigma_{1-\text{Loop}}$, we use Feynman parametrization to rewrite these expressions as
\begin{eqnarray}
    J_{a,b}^{d} = \dfrac{\Gamma\left(a+b-\frac{d}{2}\right)}{\Gamma(a)\Gamma(b)} \int_0^1 dx\, \dfrac{x^{ \frac{d}{2} -b-1}(1-x)^{b-1}}{[m^2-p^2(1-x)]^{a+b-\frac{d}{2}}}. \nonumber \\
\end{eqnarray}
Using these expression along with Eqs.~\eqref{fermion_se_l} and~\eqref{FM_1Loop}, and the $\overline{\text{MS}}$ scheme to renormalize RQED with $\dg=4$, and $\de=3$ dimensions, we obtain
\begin{eqnarray}
    F(p;\xi) &=& 1 + \dfrac{\alpha}{12\pi}\Bigg[ \dfrac{10}{3} - \ln \left( -\dfrac{p^2-m^2}{\overline{\mu}} \right) - \ln(4) \nonumber \\
    && + 2 \dfrac{m^2}{p^2} - \dfrac{m^3}{p^3} \ln(G) + 3\Bigg( \dfrac{m^3}{p^3} \ln(G) - 2 \nonumber \\
    &&- 2 \dfrac{m^2}{p^2} + \ln\left( -\dfrac{p^2-m^2}{\overline{\mu}} \right) + \ln(4) \Bigg) \Bigg] + O(\alpha^2) \,, \nonumber \\
    M(p;\xi) &=& m \Bigg\{ 1 + \dfrac{\alpha}{4\pi p^3} \Bigg[ - \dfrac{1}{3}(m^3+15m\,p^2)\ln(G) \nonumber \\
    && + \dfrac{2}{9} p \bigg( 3m^2 + 77p^2 - 24p^2\bigg( \ln\left( -\dfrac{p^2-m^2}{\overline{\mu}} \right) \nonumber \\
    && + \ln(4) \bigg) \bigg) + \xi (m^2-p^2)\left( m \ln(G)-2p \right) \Bigg] \Bigg\} \nonumber \\
    && + O(\alpha^2) \,, \label{FM_1L_final}
\end{eqnarray}
where $\overline{\mu}^2=4\pi\,\text{e}^{-\gamma_E} \,\mu^2$.

From Eqs.~\eqref{FM_LKF_LinearO} and~\eqref{FM_1L_final}, we verify that the following equations hold at linear order in $\alpha$:
\begin{eqnarray}
    F(p;\xi) - F(p;0) &=& F^{\text{LKF}}(p;\xi) - 1 \,, \nonumber \\
    M(p;\xi) - M(p;0) &=& M^{\text{LKF}}(p;\xi) - m \,.
\end{eqnarray}
Thus, as expected, the LKF approach and the one loop calculation are equivalent up to some term that would solely arise in the Landau gauge $\xi = 0$.

\bibliographystyle{apsrev4-1}

\bibliography{references.bib}

\end{document}